%% file: SymanzikStaggered.tex
\documentclass[11pt,a4paper]{article}
\usepackage{amsmath,amsfonts,amssymb,amsbsy,bbold,mathtools}
\usepackage{mathrsfs,latexsym,tikz}
\usepackage{slashed}
\usepackage{graphicx}
\usepackage{subcaption}
\usepackage{color}
\usepackage{booktabs}
\usepackage{multirow}
\usepackage{multicol}
\usepackage{alpha}
\usepackage{mathrsfs}

\usepackage{textcomp}

\usepackage{pdflscape}

\usepackage[normalem]{ulem}

\input{defs.sty}

\begin{document}

\input{title.tex}


\input{intro.tex}

\input{action.tex}

\input{basis.tex}

\input{AD.tex}

\input{concl.tex}

\appendix
\input{appa.tex}

\vskip 0.3cm

\noindent

\bibliographystyle{JHEP}

\input{SymanzikStaggered.bbl}

\end{document}

%% file: title.tex
\preprintno{%
\today
}

\title{\boldmath Logarithmic corrections to O($a^2$) effects in lattice QCD with unrooted Staggered quarks 
}

\author[ift,uam]{Nikolai~Husung}
\address[ift]{Instituto de Física Teórica UAM-CSIC, C/ Nicolás Cabrera 13-15, Universidad Autónoma de Madrid, Cantoblanco 28049 Madrid, Spain}
\address[uam]{Departamento de Física Teórica, Universidad Autónoma de Madrid, Cantoblanco 28049 Madrid, Spain}

\vspace*{-1cm}

\begin{abstract}
We derive the asymptotic lattice-spacing dependence $a^2[2b_0\gbar^2(1/a)]^{\hat{\gamma}_i}$ relevant for spectral quantities of lattice QCD, when using unrooted Staggered quarks.
Without taking any effects from matching into account we find $\min_i\hat{\gamma}_i\approx -0.273, -0.301, -0.913, -2.614$ for $\Nf=0,4,8,12$ respectively.
Common statements in the literature on the absence of mass-dimension~5 operators from the on-shell basis of the Symanzik Effective Field Theory action are being clarified for a description using strictly local tastes, here playing the role of continuum quark flavours.
Potential impact of $\ord(a)$ EOM-vanishing terms beyond spectral quantities is being discussed.
\end{abstract}

\begin{keyword}
Lattice QCD \sep Scaling \sep Effective theory \sep Staggered quarks
\end{keyword}

\maketitle

%% file: intro.tex
\section{Introduction}
A common source of systematic errors affecting all predictions from lattice QCD calculations stems from taking the continuum limit.
Even after proper renormalisation lattice artifacts, i.e., remnant regulator-dependence persists.
To remove this dependence one has to take the continuum limit $a\searrow 0$, where $a$ is the lattice spacing.

Working with numerical simulations of lattice QCD, only a limited number of finite lattice spacings $a>0$ is accessible.
Moreover going to smaller lattice spacings quickly becomes prohibitively expensive despite impressive progress on the numerical algorithms, see e.g.~\cite{Finkenrath:2023sjg} and references therein. 
Taking the continuum limit thus turns into performing continuum extrapolations involving a fairly limited number of lattice spacings, which induces a systematic error on the final predictions for continuum physics.
Since this systematic error is unavoidable some care has to be taken to properly estimate it as well as keeping it small.

The method of choice is Symanzik Effective theory (SymEFT)~\cite{Symanzik:1979ph,Symanzik:1981hc,Symanzik:1983dc,Symanzik:1983gh}, see also~\cite[p.~39ff.]{Weisz:2010nr}.
It allows to describe the lattice artifacts in terms of a continuum Effective Field Theory expanding around the desired continuum limit, here QCD.
The corresponding effective Lagrangian reads to leading order in the lattice spacing
\begin{align}
\Leff &= \L_\text{QCD}+a^2\sum_i\omega_i\opSix_i+\ord(a^3),\\
\L_\text{QCD} &= -\frac{1}{2g_0^2}\tr(F_{\mu\nu}F_{\mu\nu})+\bar{\Psi}\left\{\gamma_\mu D_\mu(A)+M\right\}\Psi,
\end{align}
where $\Psi=(\psi_1,\ldots,\psi_{\Nf})^\mathrm{T}$ is a vector of all quark flavours $\psi_i$ with $\Nf\in \{0,4,8,12\}$, $D_\mu(A)=\partial_\mu+A_\mu$ is the continuum covariant derivative with $A_\mu\in\su(\Nc)$, $F_{\mu\nu}=[D_\mu,D_\nu]$ is the field-strength tensor, and $M=\diag(m_1\mathbb{1}_{4\times 4},\ldots,m_{\Nf/4}\mathbb{1}_{4\times 4})$ are the 4-fold degenerate quark masses.\footnote{With slight abuse of notation we use here $\diag(K_1,\ldots,K_n)$ to denote block-diagonal matrices where the $K_i$ are the block matrices.}
Insisting here on having sets of four quarks that are mass-degenerate each is due to discussing unrooted Staggered quarks~\cite{Kogut:1974ag}, which have precisely this flavour structure in the continuum limit.
We will introduce the lattice formulation in some detail in \sect{sec:action}.
The operators $\opSix_i$ form a minimal basis only constrained by the symmetries of the lattice action and of course their canonical mass-dimension.
Here $\omega_i$ are the bare matching coefficients which for now are kept free and can be chosen according to the details of the lattice formulation, e.g., incorporating any kind of improvement.
In \sect{sec:basis} we derive the minimal on-shell operator basis of the SymEFT action compatible with the symmetries of the Staggered lattice formulation.

In \sect{sec:AD} we eventually obtain the leading asymptotic lattice spacing dependences $a^2[2b_0\gbar^2(1/a)]^{\hat{\gamma}_i}$ with $2b_0\hat{\gamma}_i=\gamma_0^\base$ and $48\pi^2b_0=11\Nc-2\Nf$.
As has been discussed in more detail in~\cite{Husung:2022kvi}, the central ingredient is the Jordan normal form of the 1-loop anomalous dimension matrix $\gamma_0^\base$ of the minimal on-shell operator basis.

%% file: action.tex
\section{The Staggered-quark action}\label{sec:action}
The lattice action for Staggered quarks is usually derived from so called \textit{naive} quarks~\cite{Wilson:1974}, here chosen to be massless,
\begin{equation}
S_\mathrm{F}^\mathrm{naive}=a^4\sum_x\bar{\psi}(x)\frac{\gamma_\mu}{2}\left[\nabla_\mu+\nabla_\mu^*\right]\psi(x),\label{eq:naiveWilson}
\end{equation}
where $\psi$ is a single colour-charged spinor, $\gamma_\mu$ are the hermitian $\gamma$-matrices, and the covariant forward and backward derivatives are defined as
\begin{align}
\nabla_\mu \psi(x)&=\frac{U(x,\mu)\psi(x+a\hat{\mu})-\psi(x)}{a},\\
\nabla_\mu^*\psi(x)&=\frac{\psi(x)-U^\dagger(x-a\hat{\mu},\mu)\psi(x-a\hat{\mu})}{a},
\end{align}
where $U(x,\mu) \in $~SU($N$) are the link variables connecting $x+a\hat\mu$ and $x$.
This lattice action implements all the desired continuum flavour symmetries well, but introduces 15 additional poles to the free quark propagator in the Brillouin zone $ap_\mu\in]-\pi,\pi]$ or equivalently zeroes to the free lattice Dirac operator
\begin{equation}
a\hat{\tilde{D}}_\mathrm{free}^\mathrm{naive}(p)=i\sum_\mu\gamma_\mu \sin(ap_\mu)\,.
\end{equation}
Those poles result in the simulation of $2^D$ flavours, i.e., here 16 in total rather than the expected single flavour.
This is a realisation of the well-known Nielsen-Ninomiya no-go theorem~\cite{Nielsen:1980rz, Nielsen:1981xu, Nielsen:1981hk} regarding the impossibility of an ultra-local lattice action with exact (continuum) chiral symmetry without such additional poles.
Ways around this problem typically eliminate the additional poles, see e.g.~\cite{Wilson:1975id, Ginsparg:1981bj, Kaplan:1992bt, Furman:1994ky, Neuberger:1997fp, Neuberger:1998wv}, with varying impact on the flavour-symmetries still realised on the lattice and vastly different computational cost in Monte Carlo simulations when, e.g., applying the lattice Dirac operator.

We will focus here on yet another solution to this problem, namely the Staggered quark action~\cite{Kogut:1974ag} also known as Kogut-Susskind (KS) quarks.
It can be derived from \eq{eq:naiveWilson} using the \textit{Staggered transformation}
\begin{equation}
\bar{\psi}(x)=\bar{\chi}(x)\Bigg\{\prod_{\mu=0}^3 \gamma_\mu^{x_\mu/a}\Bigg\}^\dagger,\quad \psi(x)=\prod_{\mu=0}^3 \gamma_\mu^{x_\mu/a}\chi(x),
\end{equation}
where the product multiplies from the right.
The action becomes diagonal in the transformed spinor components and one simply discards three of the four components.\footnote{With abuse of notation we will denote the single remaining spinor component also with $\chi$.}
This reduces the number of flavours to four and leads to the action of (unrooted) Staggered quarks
\begin{equation}
S_\mathrm{F}[\bar{\chi},\chi]=a^4\sum_{x}\bar{\chi}(x)\frac{\eta_\mu(x)}{2}\left[\nabla_\mu+\nabla_\mu^*\right]\chi(x)\,,\label{eq:staggeredAction}
\end{equation}
where
\begin{equation}
\eta_\mu(x)=(-1)^{\sum\limits_{\nu<\mu}x_\nu/a}.
\end{equation}
Working in this one-component representation~(1CR) makes it very difficult to identify the actual flavours and what remains of the continuum symmetries.
We thus employ yet another transformation to identify so called \textit{tastes}, which play the role of flavours in our theory.

In principle there are various ways to define those tastes.
Here we are only limited by the requirement of having a \emph{local} lattice action to apply Symanzik Effective theory.\footnote{Lattice artifacts (and UV divergences) are short-distance effects and therefore require a description that is local.}
Despite having advantageous flavour-symmetries in the free theory, the so-called momentum-space representation~(MR), see e.g.~\cite{Sharatchandra:1981si,Golterman:1984cy}, is immediately ruled out due to its non-locality~\cite{Kluberg-Stern:1983lmr,Daniel:1986zm,Jolicoeur:1986ek}.

We follow here along the lines of \cite{Verstegen:1985kt} to introduce such a taste-representation~(TR) that is strictly local in spacetime.
First we join the 16 sites of a hypercube at reference point $y$ and label the corners by the coordinate $\xi$, i.e.,
\begin{equation}
x=y+a\xi\,,\quad x\in a\mathbb{Z}^4\,,\quad \xi\in\{0,1\}^4\,,\quad y\in 2a\mathbb{Z}^4\,.
\end{equation} 
Within each of these hypercubes we can then define tastes
\begin{equation}\label{eq:spacetimeTastes}
   \tastebar^\beta_{\alpha}(y)=\frac{1}{4}\sum_\xi\bar{\chi}(y+a\xi)\tastedef^\dagger_{\beta\alpha}(y,\xi)\,,\quad \taste^\beta_{\alpha}(y)=\frac{1}{4}\sum_\xi \tastedef_{\alpha\beta}(y,\xi)\chi(y+a\xi)\,,
\end{equation}
where we introduced the shorthand
\begin{align}
   \tastedef_{\alpha\beta}(y,\xi)&=\frac{1}{2}W(y,\xi)\Gamma_{\alpha \beta}(y+a\xi),\\
   W(y,\xi)&=\prod_\mu U^{\xi_\mu}\left(y+a{\textstyle\sum_{\nu<\mu}}\xi_\nu\hat{\nu},\mu\right),&\Gamma(z)&=\gamma_0^{z_0/a}\gamma_1^{z_1/a}\gamma_2^{z_2/a}\gamma_3^{z_3/a}.\nonumber
\end{align}
This prescription makes use of the completeness relations of Dirac gamma matrices, see also~\cite[p.~64f.]{Rothe:1992nt}.
The Wilson line ensures gauge covariance.
There is quite some freedom on how to connect the corners of the hypercube to the reference point via a Wilson line for all the combinations of spin and flavour while keeping $\tastedef$ unitary.
Tastes are labelled as $\Phi$ to avoid later confusion with the continuum quarks.
It is important to realise that this change of basis leaves the path integral measure invariant up to normalisation, i.e.,
\begin{equation}
\pD \bar{\chi}\,\pD\chi\propto\pD\tastebar\,\pD\taste.
\end{equation}
Otherwise we would have to deal with a contribution of the measure to the tastes' lattice action.
Clearly the tastes carry twice the number of ``spinor indices'' $\alpha$ and $\beta$ than one would naively expect.
To understand this apparent mismatch, we need to remind ourselves that we are looking for four flavours rather than one and we identify $\beta$ as a taste index labelling our ``flavour''.
The resulting lattice action then reads in the free theory
\begin{equation}
S_\mathrm{F}[\tastebar,\taste]=(2a)^4\sum_{\mathclap{y\in (2a\mathbb{Z})^4}}\tastebar(y)\left\{\frac{\gamma_\mu\otimes \unity }{2}\left[\hat\partial_\mu+\hat\partial_\mu^*\right]+\frac{\gamma_5\otimes \tau_\mu\tau_5}{2}\left[\hat\partial_\mu-\hat\partial_\mu^*\right]\right\}\taste(y)\,,\label{eq:Staste}
\end{equation}
where the sum now runs only over the hypercube indices $y$ and $\tau_\mu=\gamma_\mu^T$ acts in taste-space rather than spinor-space.
Notice that the term in \eq{eq:Staste} carrying $\gamma_5\otimes\tau_\mu\tau_5$ plays a role very similar to a Wilson term cancelling doublers.
The forward and backward derivatives in \eq{eq:Staste} span two lattice spacings
\begin{align}
\hat\partial_\mu \taste(y)&=\frac{\taste(y+2a\hat{\mu})-\taste(y)}{2a}\,,&
\hat\partial_\mu^* \taste(y)&=\frac{\taste(y)-\taste(y-2a\hat{\mu})}{2a}\,.
\end{align}
Other choices for the reference point $y$ and orientations of the hypercube are expected to be equivalent, thus amounting to different unitary transformations in field space keeping the symmetries of the 1CR intact.
For a full list of those symmetries, see e.g.~\cite{Golterman:1984cy,Verstegen:1985kt} and the collection thereof in \app{sec:1CRsymmetries} with some important remarks.
We only give here the action in the free theory as its form in the fully interacting theory becomes rather involved within our conventions while providing the reader with no more insight.

Notice that the free action is also the natural discretisation of K\"ahler-Dirac~(KD) fermions~\cite{Kahler2011,Graf:1978kr,Rabin:1981qj}.
This should come as no surprise given that KD fermions in flat 4-dimensional spacetime are equivalent to four mass-degenerate quarks.
Depending on our choice of how to couple the quarks to the gauge sector then sets KS and KD fermions apart.
For KD fermions it would be natural to introduce the gauge links directly in the TR.
As it turns out~\cite{Mitra:1983bi}, such a choice breaks Shift-symmetry of the free 1CR~\eq{eq:1CRshift} and thus allows for additive mass-renormalisation.
Instead we choose KS fermions as introduced before and commonly used in the literature.

%% file: basis.tex
\section{Symmetries of Staggered quarks and the minimal SymEFT operator basis}\label{sec:basis}
To find the minimal basis of operators up to mass-dimension~6 and thus $\ord(a^2)$ in the lattice spacing, we first need to work out the symmetry constraints imposed by the chosen Staggered quarks lattice discretisation.
As we established in the previous section, we will use here the TR from \eq{eq:spacetimeTastes} to ensure a local formulation of the lattice action.

It is imperative to remember that we have to use the symmetry constraints of the fully interacting theory.
Any argument based on the free theory in a way that gets broken by the introduction of gauge links must be avoided at all costs.
We must stress this again as it is well known~\cite{Mitra:1983bi,Golterman:1984cy} that the details on how the gauge links are introduced into the free theory decide whether symmetries, like e.g.~Shift-symmetry, survive at all or need to be modified.

To easily identify the symmetry transformations realised in the TR, we start from the counterparts in the interacting 1CR, see \app{sec:1CRsymmetries}, and map them to our choice of tastes according to \eq{eq:spacetimeTastes}.
Having the transformations in terms of the TR, finally allows us to make the connection to the (continuum notion of) flavour in the SymEFT.
To summarize, the following symmetry constraints of the lattice action must be reproduced by our SymEFT operator basis:
\begin{subequations}\label{eq:SymmetryTransfs}
\begin{itemize}
\item SU($\Nc$) gauge symmetry,
\item U(1)${}_\mathrm{B}$ flavour symmetry,
\item Remnant of chiral symmetry.
Analogously to conventional chiral symmetry we may introduce the shorthands
\begin{align}
\tastebar_\mathrm{R}&=\tastebar\frac{1-\gamma_5\otimes\tau_5}{2}\,,&\taste_\mathrm{R}&=\frac{1+\gamma_5\otimes\tau_5}{2}\taste\,,\nonumber\\\tastebar_\mathrm{L}&=\tastebar\frac{1+\gamma_5\otimes\tau_5}{2}\,,&
\taste_\mathrm{L}&=\frac{1-\gamma_5\otimes\tau_5}{2}\taste\,.
\end{align}
The massless lattice action written in this form is then invariant under
\begin{align}
\tastebar_\mathrm{R}&\rightarrow\tastebar_\mathrm{R}e^{-i\varphi_\mathrm{R}},& \taste_\mathrm{R}&\rightarrow e^{i\varphi_\mathrm{R}}\taste_\mathrm{R},\nonumber\\
\tastebar_\mathrm{L}&\rightarrow\tastebar_\mathrm{L}e^{-i\varphi_\mathrm{L}},&\taste_\mathrm{L}&\rightarrow e^{i\varphi_\mathrm{L}}\taste_\mathrm{L},
&\varphi_\mathrm{L,R}&\in\mathbb{R}.\label{eq:LRsymm}
\end{align}
\item \emph{Modified} charge conjugation
\begin{align}
\tastebar(y)&\rightarrow-\taste^T(y)C\otimes (C^{-1})^T,\quad \taste(y)\rightarrow C^{-1}\otimes C^T\tastebar^T(y),\nonumber\\
U(x,\mu)&\rightarrow U^*(x,\mu),\quad C\gamma_\mu C^{-1}=-\gamma_\mu^T.
\end{align}
\item[$\circ$] \emph{Modified} Euclidean reflections~\cite{Verstegen:1985kt} in $\hat{\mu}$ direction
\begin{align}
\tastebar(y)&\rightarrow \tastebar(y-2y_\mu\hat{\mu})\gamma_5\gamma_\mu\otimes \tau_5\left\{1+a^2\ldots\right\},\nonumber\\
\taste(y)&\rightarrow\gamma_\mu\gamma_5\otimes\tau_5\left\{1+a^2\ldots\right\}\taste(y-2y_\mu\hat{\mu})\,,\nonumber\\
U(x,\nu)&\rightarrow\begin{cases}
U^\dagger(x-(2x_\mu+a)\hat{\mu},\mu) & \mu=\nu \\
U(x-2x_\mu \hat{\mu},\mu)U(x-(2x_\mu-a)\hat{\mu},\nu)U^\dagger(x-2x_\mu\hat{\mu}+a\hat{\nu},\mu) & \text{else}
\end{cases}.\label{eq:modReflections}
\end{align}
\item[$\circ$] \emph{Modified} discrete rotations~\cite{Mitra:1983bi,Verstegen:1985kt} of $90^\circ$ in any $\rho$-$\sigma$-plane
\begin{align}
\tastebar(y)&\rightarrow\frac{1}{2}\tastebar(R^{-1}y)(\unity +\gamma_\rho\gamma_\sigma)\otimes (\tau_\sigma-\tau_\rho)\{1+a^2\ldots\},\nonumber\\
\taste(y)&\rightarrow\frac{1}{2}(\unity -\gamma_\rho\gamma_\sigma)\otimes (\tau_\sigma-\tau_\rho)\{1+a^2\ldots\}\taste(R^{-1}y),\nonumber\\
U(x,\nu)&\rightarrow\begin{cases}
U^\dagger(R^{-1}x-a\hat{\sigma},\sigma) & \nu=\rho\\
U(R^{-1}x,\sigma)U(R^{-1}x+a\hat{\sigma},\rho)U^\dagger(R^{-1}x+a\hat{\rho},\sigma) & \nu=\sigma\\
U(R^{-1}x,\sigma)U(R^{-1}x+a\hat{\sigma},\nu)U^\dagger(R^{-1}x+a\hat{\nu},\sigma) & \text{else}
\end{cases}\nonumber\\
(R^{-1}x)_\rho&=x_\sigma,\quad (R^{-1}x)_\sigma=-x_\rho,\quad (R^{-1}x)_{\mu\neq \rho,\sigma}=x_\mu,
\end{align}
with rotation matrix $R$ acting on vectors in Euclidean spacetime.
\item[$\circ$] Shift-symmetry by a single lattice spacing in direction $\hat{\mu}$ amounting to a discrete flavour rotation and field-redefinition
\begin{align}
\bar\taste(y)&\rightarrow \bar\taste(y)\unity\otimes\tau_\mu \left\{1+2a\hat{P}_-^{(\mu)}\hat{\overline{\nabla}}{}_\mu^{\dagger}+a^2\ldots\right\},\nonumber\\
\taste(y)&\rightarrow \unity\otimes\tau_\mu \left\{1+2a\hat{P}_+^{(\mu)}\hat{\overline{\nabla}}_\mu+a^2\ldots\right\}\taste(y),\nonumber\\
U(x,\nu) &\rightarrow U(x,\mu)U(x+a\hat{\mu},\nu)U^\dagger(x+a\hat{\nu},\mu),\label{eq:ShiftTaste}
\end{align}
where $\hat{\overline{\nabla}}_\mu$ involves a $\hat\mu$-dependent fat link and we introduced the projectors
\begin{equation}
\hat{P}_\pm^{(\mu)}=\frac{1\pm\gamma_\mu\gamma_5\otimes\tau_\mu\tau_5}{2}.
\end{equation}
\end{itemize}
\end{subequations}
Filled ($\bullet$) or open ($\circ$) symbols highlight symmetries that are exact counterparts of their continuum-theory variant or require field-redefinitions respectively of both $\tastebar$ and $\taste$ \emph{in the lattice theory}.
For an explicit example of such a field-redefinition see \app{sec:RedefExample}.
To be clear, we simply omit here lattice discretisations involved in the field-redefinitions of canonical mass-dimension~2, like a lattice field-strength, and beyond for readability and no expansion in the lattice spacing takes place on the lattice theory side.
The full expressions of these field-redefinitions can be obtained with help of the supplemented \texttt{Mathematica} notebook.

The actual impact of the full field-redefinitions on the SymEFT description can be deduced from the lattice transformation properties as follows.
Aside from Shift-symmetry, all field-redefinitions stay within the same hypercube used to define the tastes.
In those cases, we may introduce a \emph{local} phase $\taste(y)\rightarrow e^{i\varphi(y)}\taste(y)$ prior to performing the full symmetry transformation and absorb the phase again a-posteriori.
Meanwhile, due to the Shift-symmetry transformation spanning two (coarse) lattice points the local phases will no longer cancel.
Thinking of an off-shell matching strategy for the SymEFT involving vertex-functions of quark fields, these transformation properties imply that covariant derivatives acting on the quark fields are only allowed for the field-redefinition of Shift-symmetry.
All other symmetry transformations may only be parametrised in terms of the identity and (higher-order covariant derivatives of) field-strengths, which then implies that those field-redefinitions may have an impact at the earliest at $\ord(a^2)$.
This is also compatible with the observation that the field-redefinitions acting on the lattice quark field gauge-transform in the adjoint representation for all symmetries but Shift-symmetry.
The reasoning used here should hold to all orders in the lattice spacing within SymEFT.

\emph{Modified} Euclidean reflection in the $\hat{0}$ direction is, for our specific choice of the TR, special as it corresponds to time reflection combined with a $\tau_5$-flavour rotation in its canonical continuum form (CCF).
All \emph{modified} reflections in any of the other directions require a field-redefinition in the lattice theory, where the $\hat{1}$-direction is being used as an example in \app{sec:RedefExample}.
Even in the free theory, Euclidean reflections (and therefore parity as well as time reversal) are only realised in combination with an additional taste-rotation $\tau_5$ due to the ``irrelevant'' term that vanishes in the continuum limit.
For the same reason charge conjugation and discrete rotations must be combined with a transformation in taste space.\footnote{At least Euclidean reflections can be restored to their original continuum form in the free theory, by changing the definition of the tastes as suggested in~\cite{Mitra:1983bi}, but we use here a slightly different convention
\begin{equation}
\tastebar=\tastebar'\left\{\unity\otimes Q_++i\gamma_5\otimes Q_-\right\},\quad
\taste=\left\{\unity\otimes Q_++i\gamma_5\otimes Q_-\right\}\taste',\quad 
Q_\pm=\frac{\unity\pm\tau_5}{2}\,,
\end{equation}
to keep \emph{modified} charge conjugation as it was.
We stick to the more common variant without this redefinition.}
Only Shift-symmetry requires a field-redefinition already in the free theory, where then only the forward difference survives $\hat{\overline{\nabla}}_\mu\rightarrow\hat{\partial}_\mu$.

\subsection{Implications of symmetries involving field-redefinitions}\label{sec:symmRedef}
Readers familiar with the formulation of Staggered quarks may have noticed the peculiar structure of \emph{modified} Euclidean reflections, \emph{modified} discrete rotations, and Shift-symmetry in \eqs{eq:SymmetryTransfs}.
All the aforementioned symmetries have in common that they deviate from the CCF of the symmetry and require a field-redefinition to restore the original hypercubic construction of the TR in the interacting theory at finite lattice spacing.
Those more involved symmetries are still sufficient to protect the theory from power divergences.
Also the CCF of the symmetry should clearly get restored in the continuum limit.
Now we only need to understand what effect such field-redefinitions have on the allowed form of lattice artifacts.
Notice that by construction, the measure of the lattice path integral stays invariant under those field-redefinitions.

At linear order, a field-redefinition will only give rise to EOM-vanishing operators in the SymEFT.
Consequently, a symmetry that requires a field-redefinition in the lattice theory will give rise to EOM-vanishing operators in the SymEFT violating this specific (continuum) symmetry.
Those EOM-vanishing operators \emph{must} be present also in the SymEFT when using an off-shell matching strategy to capture the lattice-symmetry's interplay of different orders in the lattice spacing.
Conceptually, those EOM-vanishing operators can be absorbed via adjusting the matching to the fundamental quark fields in the SymEFT, see the discussion of this for Wilson-type quarks \cite{Husung:localFields}.
Thus, we can always adjust the matching conditions of the SymEFT to absorb any deviation of the action from the exact continuum symmetry at fixed order in the lattice spacing.
For the on-shell basis of the SymEFT action this will impact the matching beyond $\ord(a)$ due to Shift-symmetry as well as beyond $\ord(a^3)$ due to \emph{modified} rotations and \emph{modified} Euclidean reflections, i.e., the quadratic pieces the changes of matching conditions.
For a sketch of a proof that symmetry transformations involving field-redefinitions can be restored to their CCF order by order in the lattice spacing, see \app{sec:symmEOMrestoration}.

However, any such change of matching condition for the fundamental fields has to be propagated into composite local fields like, e.g., currents to account for contact terms of the EOM-vanishing operators with the local fields present prior to changing the matching conditions.
The impact on lattice artifacts of composite local fields when restoring Shift-symmetry to a conventional discrete flavour rotation already sets in at $\ord(a)$ but the discrete flavour-symmetry of the continuum theory
\begin{equation}
\ctastebar(y)\rightarrow \ctastebar(y)\tau_5,\quad \ctaste(y)\rightarrow\tau_5\ctaste(y), \quad \{\tau_5\}\subset\mathrm{SU}(4)\label{eq:tau5Symmetry}
\end{equation}
should ensure \emph{automatic $\ord(a)$ improvement} of any matrix element or correlator with a non-trivial continuum limit.
This is very similar to the automatic $\ord(a)$ improvement of twisted mass QCD~\cite{Frezzotti:1999vv,Frezzotti:2000nk,Frezzotti:2005gi} at maximal twist~\cite{Aoki:2006gh,Sint:2007ug}.
Here and in the following $\ctaste$ denotes the flavour-vector of the continuum theory containing the four mass-degenerate flavours corresponding to one set of tastes of the lattice theory.

\subsection{Minimal operator basis to $\ord(a^2)$}
In contrast to the continuum QCD action, the flavour symmetries are severely broken and within each set of tastes taste-changing interactions occur at finite lattice spacing.
This is only expected to be cured in the limit $a\searrow 0$, but will obviously be reflected in the minimal operator basis of SymEFT describing the lattice artifacts at finite lattice spacing $a>0$.
As in earlier publications, we can discard any operators vanishing by the continuum equations of motion~(EOM)~\cite{Luscher:1996sc}
\begin{equation}
[D_\mu(A), F_{\mu\nu}]=T^a\bare{g}^2\ctastebar\gamma_\nu T^a\ctaste,\quad \gamma_\mu D_\mu(A) \ctaste=-M\ctaste,\quad \ctastebar \cev D_\mu(A) \gamma_\mu=\ctastebar M,\label{eq:EOMs}
\end{equation}
to obtain a minimal on-shell operator basis for our SymEFT action.
Here $T^a$ denotes a generator of the $\su(\Nc)$ colour algebra and $\ctaste$ is the vector containing all flavours.
In practice, one should keep track of such operators as the field-redefinitions involved to eliminate EOM-vanishing operators will impact any SymEFT analysis of local composite fields as laid out in detail in~\cite{Capitani:1999ay,Capitani:2000xi,Husung:localFields} and in general subleading corrections in the lattice spacing.

\subsubsection{Mass-dimension~5}\label{sec:minimalOnshellBasis5}
Unfortunately the symmetry constraints from~\eqs{eq:SymmetryTransfs} are insufficient to exclude EOM-vanishing operators of mass-dimension~5 while there are no operators allowed for the minimal on-shell basis of the SymEFT action.
This is in slight contradiction to what has been found in cf.~\cite{Luo:1996vt} and we find
\begin{align}
\opFiveE[;1]&=\frac{1}{2}\sum_\nu\ctastebar\left\{\cev{\Dslash}\cev{D}_\nu\gamma_\nu\gamma_5\otimes\tau_\nu+\gamma_\nu\gamma_5\otimes \tau_\nu D_\nu\Dslash\right\}\ctaste,\nonumber\\
\opFiveE[;2]&=\frac{1}{2}\sum_\nu\ctastebar\left\{\cev{\Dslash}M\gamma_5\otimes\tau_\nu+\gamma_5\otimes\tau_\nu M\Dslash\right\}\ctaste.\label{eq:opFive}
\end{align}
We use here and in the following the sloppy shorthands $\Dslash \Psi=(\gamma_\kappa\otimes\unity  D_\kappa + M)\Psi$ and $\bar{\Psi}\cev{\Dslash}=\bar{\Psi}(\gamma_\kappa\otimes\unity  \cev{D}_\kappa - M)$.
As mentioned earlier such EOM-vanishing terms can be removed order by order to restore Shift-symmetry of the SymEFT action to its canonical form of simple discrete flavour rotations $\tau_\mu$.
All EOM-vanishing operators found here transform odd under the discrete flavour symmetry in \eq{eq:tau5Symmetry} of continuum QCD.
This ensures automatic $\ord(a)$ improvement even for contact terms of those operators with any local fields as long as the considered continuum limit does not trivially vanish.
Nonetheless this will have an impact on $\ord(a^2)$ lattice artifacts of correlators and matrix elements and matching coefficients of the on-shell basis of the SymEFT action.

\subsubsection{Mass-dimension~6}\label{sec:minimalOnshellBasis6}
Due to the reduced flavour symmetries compared to GW quarks a plethora of operators becomes relevant at mass-dimension~6 besides those already known from Ginsparg-Wilson quarks~\cite{Sheikholeslami:1985ij,Husung:2022kvi}.
For the latter a suitable minimal on-shell basis is
\begin{align}
\opSix_{1}&=\frac{1}{\bare{g}^2}\tr([D_\mu, F_{\nu\rho}]\,[D_\mu, F_{\nu\rho}])
\,,&
\opSix_{2}&=\frac{1}{\bare{g}^2}\sum\limits_{\mu}\tr([D_\mu, F_{\mu\nu}]\,[D_\mu, F_{\mu\nu}])\,,&\nonumber\\
\opSix_{3}&=\sum_\mu\bar\ctaste\gamma_\mu\otimes\unity  D_\mu^3\ctaste,&
\opSix_{4}&=\bare{g}^2(\bar\ctaste\gamma_\mu\otimes\unity \ctaste)^2,\nonumber\\
\opSix_{5}&=\bare{g}^2(\bar\ctaste\gamma_\mu\gamma_5\otimes\unity \ctaste)^2,&
\opSix_{6}&=\bare{g}^2(\bar\ctaste\gamma_\mu\otimes\unity  T^a\ctaste)^2,\nonumber\\
\opSix_{7}&=\bare{g}^2(\bar\ctaste\gamma_\mu\gamma_5\otimes\unity T^a\ctaste)^2,&
\opSix_{8}&=\frac{i}{4}\ctastebar M\gamma_{\mu\nu}\otimes\unity F_{\mu\nu}\ctaste,\nonumber\\
\opSix_{9}&=\ctastebar M^3\ctaste,&
\opSix_{10}&=\tr(M^2)\ctastebar M\ctaste.\label{eq:4fermionChiral}
\end{align}
By convention, we use Einstein's summation convention and explicit sums are only introduced to indicate O(4) symmetry breaking. 
Any seemingly free indices in the 4-quark operators are to be summed over with the one from the second quark-anti-quark pair.
The extension of the minimal on-shell basis only compatible with the remnant of chiral symmetry reads
\begingroup\allowdisplaybreaks
\begin{align}
\opSix_{11}&=\bare{g}^2(\ctastebar\unity \otimes \tau_\mu\ctaste)^2,&
\opSix_{12}&=\bare{g}^2(\ctastebar\unity \otimes \tau_\mu\tau_5\ctaste)^2,\nonumber\\
\opSix_{13}&=\bare{g}^2(\ctastebar\unity \otimes \tau_\mu T^a\ctaste)^2,&
\opSix_{14}&=\bare{g}^2(\ctastebar\unity \otimes \tau_\mu\tau_5 T^a\ctaste)^2,\nonumber\\
\opSix_{15}&=\bare{g}^2(\ctastebar\gamma_5\otimes \tau_\mu \ctaste)^2,&
\opSix_{16}&=\bare{g}^2(\ctastebar\gamma_5\otimes \tau_\mu\tau_5 \ctaste)^2,\nonumber\\
\opSix_{17}&=\bare{g}^2(\ctastebar\gamma_5\otimes \tau_\mu T^a\ctaste)^2,&
\opSix_{18}&=\bare{g}^2(\ctastebar\gamma_5\otimes \tau_\mu\tau_5 T^a\ctaste)^2,\nonumber\\
\opSix_{19}&=\bare{g}^2(\ctastebar\gamma_\mu\otimes \tau_5 \ctaste)^2,&
\opSix_{20}&=\bare{g}^2(\ctastebar\gamma_\mu\gamma_5\otimes \tau_5 \ctaste)^2,\nonumber\\
\opSix_{21}&=\bare{g}^2(\ctastebar\gamma_\mu\otimes \tau_5 T^a\ctaste)^2,&
\opSix_{22}&=\bare{g}^2(\ctastebar\gamma_\mu\gamma_5\otimes \tau_5 T^a\ctaste)^2,\nonumber\\
\opSix_{23}&=\bare{g}^2(\ctastebar\gamma_{\mu\nu}\otimes \tau_\rho \ctaste)^2,&
\opSix_{24}&=\bare{g}^2(\ctastebar\gamma_{\mu\nu}\otimes \tau_\rho\tau_5 \ctaste)^2,\nonumber\\
\opSix_{25}&=\bare{g}^2(\ctastebar\gamma_{\mu\nu}\otimes \tau_\rho T^a\ctaste)^2,&
\opSix_{26}&=\bare{g}^2(\ctastebar\gamma_{\mu\nu}\otimes \tau_\rho\tau_5 T^a\ctaste)^2,\nonumber\\
\opSix_{27}&=\bare{g}^2(\ctastebar\gamma_\mu\otimes \tau_{\nu\rho} \ctaste)^2,&
\opSix_{28}&=\bare{g}^2(\ctastebar\gamma_\mu\gamma_5\otimes \tau_{\nu\rho} \ctaste)^2,\nonumber\\
\opSix_{29}&=\bare{g}^2(\ctastebar\gamma_\mu\otimes \tau_{\nu\rho} T^a\ctaste)^2,&
\opSix_{30}&=\bare{g}^2(\ctastebar\gamma_\mu\gamma_5\otimes \tau_{\nu\rho} T^a\ctaste)^2,\nonumber\\
\opSix_{31}&=\bare{g}^2\sum_\mu(\ctastebar\gamma_{\mu\nu}\otimes \tau_\mu \ctaste)^2,&
\opSix_{32}&=\bare{g}^2\sum_\mu(\ctastebar\gamma_{\mu\nu}\otimes \tau_\mu\tau_5 \ctaste)^2,\nonumber\\
\opSix_{33}&=\bare{g}^2\sum_\mu(\ctastebar\gamma_{\mu\nu}\otimes \tau_\mu T^a\ctaste)^2,&
\opSix_{34}&=\bare{g}^2\sum_\mu(\ctastebar\gamma_{\mu\nu}\otimes \tau_\mu\tau_5 T^a\ctaste)^2,\nonumber\\
\opSix_{35}&=\bare{g}^2\sum_\mu(\ctastebar\gamma_\mu\otimes \tau_{\mu\nu} \ctaste)^2,&
\opSix_{36}&=\bare{g}^2\sum_\mu(\ctastebar\gamma_\mu\gamma_5\otimes \tau_{\mu\nu} \ctaste)^2,\nonumber\\
\opSix_{37}&=\bare{g}^2\sum_\mu(\ctastebar\gamma_\mu\otimes \tau_{\mu\nu} T^a\ctaste)^2,&
\opSix_{38}&=\bare{g}^2\sum_\mu(\ctastebar\gamma_\mu\gamma_5\otimes \tau_{\mu\nu} T^a\ctaste)^2.\label{eq:4fermionStaggered}
\end{align}
\endgroup
In agreement with \cite{Follana:2006rc} we find 28 linearly-independent 4-quark operators parametrising all taste-breaking effects at $\ord(a^2)$.
In total we thus find four more operators than have been found in~\cite{Lee:1999zxa} -- we checked this twice where the second iteration relied on an automated script.\footnote{The derivation of the minimal operator bases for local fields and different lattice actions is the primary bottleneck of the SymEFT analysis at this point.
An automation has been implemented and will be made publicly available in the near future.}
For completeness and use in our off-shell renormalisation conditions we list the minimal basis of EOM-vanishing operators of mass-dimension~6 in appendix~\ref{sec:EOMops} that are compatible with the symmetries of the on-shell basis.

The generalisation to multiple sets of tastes simply requires the replacement $\tau\rightarrow \diag(\tau,\ldots,\tau)$ since we have to do any shift on the lattice in all tastes simultaneously.
Furthermore, for a fully mass-degenerate setup we may still rotate among the $n$th flavour of each set, i.e., have an $\text{SU}(N_\mathrm{sets})^4$ flavour symmetry, where $N_\mathrm{sets}$ denotes the number of sets in the theory each containing 4 tastes that are connected via Shift-symmetry, remnant chiral, \emph{modified} Euclidean reflections, and \emph{modified} rotations as listed in \eqs{eq:SymmetryTransfs}.

%% file: AD.tex
\section{One-loop anomalous dimensions}\label{sec:AD}
The entire strategy to extract the 1-loop anomalous dimension matrix for our on-shell basis has been laid out in \cite{Husung:2022kvi}.
It can briefly be summarised by the following ingredients:
\begin{itemize}
\item Background-field gauge~\cite{DeWitt:1967ub,KlubergStern:1974xv,Abbott:1980hw,Luscher:1995vs}, which ensures that only gauge-invariant operators are allowed as counter-terms during off-shell renormalisation contrary to conventional gauge-fixing~\cite{Joglekar:1975nu,Collins:1994ee}.
\item Dimensional regularisation~\cite{tHooft:1972tcz,tHooft:1973mfk}, where all calculations take place in $D=4-2\epsilon$ dimensions.
\item Off-shell renormalisation at the level of vertex-functions with single operator insertions in the modified minimal subtraction scheme ($\MSbar{}$)~\cite{tHooft:1972tcz,tHooft:1973mfk,Bardeen:1978yd}.
\end{itemize}
With these ingredients we can derive the anomalous dimension matrix in the $\MSbar{}$ renormalisation scheme from
\begin{equation}
\mu\frac{\rmd \op_{i;\MSbar}(\mu)}{\rmd\mu}=\gamma_{ij}^\op(\gbar^2(\mu))\op_{j;\MSbar}(\mu)\,.
\end{equation}
As mentioned before, the on-shell operator basis contains a subset that is chirally symmetric from which we can infer the following structure for the anomalous dimension matrix of the on-shell basis
\begin{equation}
\gamma^\op=\begin{pmatrix}
\gamma^{m} & 0 & 0 \\[6pt]
\gamma^{\text{L}|\text{R},m} & \gamma^{\text{L}|\text{R}} & 0 \\[6pt]
\gamma^{\text{stag},m} & \gamma^{\text{stag},\text{L}|\text{R}} & \gamma^\text{stag}
\end{pmatrix},
\end{equation}
where the block matrices $\gamma^{\text{L}|\text{R}}$, $\gamma^{\text{L}|\text{R},m}$, and $\gamma^{m}$ are the same that one finds for GW quarks to $\ord(a^2)$.
We checked explicitly that this is true for our \texttt{FORM} scripts after being adapted to taste-space matrices.
The ordering of the operator basis is as in \eqs{eq:4fermionChiral} and \eqref{eq:4fermionStaggered} in their respective subblocks.
Beyond $\ord(a^2)$ the massive basis will no longer be free of taste-breaking effects.

We only give here the genuinely new block matrices and refer the reader to \cite{Husung:2022kvi} for $\gamma^{\text{L}|\text{R}}$, $\gamma^{\text{L}|\text{R},m}$, and $\gamma^{m}$ as well as an in-depth introduction into the notation being used here.
Restricting ourselves to the 1-loop mixing block matrices
\begin{equation}
\gamma^X(\gbar^2)=-\gamma_0^X\gbar^2+\ord(\gbar^4),
\end{equation}
where $X$ labels the appropriate block, the new block matrices read
\begin{align}
(4\pi)^2\gamma_0^\text{stag} &=\begin{pmatrix}
 A & 0 & 0 & B & 0 & 0 & 0 \\
 0 & A & 0 & B & 0 & 0 & 0 \\
 0 & 0 & C & 0 & 0 & 0 & 0 \\
24B&24B& 0 & D & 0 & 0 & 0 \\
 0 & 0 & 0 & 0 & C & 0 & 0 \\
 6B& 6B& 0 & E & 0 & F & 0 \\
 0 & 0 & 0 & 0 & G & 0 & H
\end{pmatrix},\nonumber\\
(4\pi)^2\gamma_0^{\text{stag},\text{L}|\text{R}} &=\begin{pmatrix}
 I \\
-I \\
 J \\
 0 \\
12J\\
 0 \\
 3J
\end{pmatrix},\quad
(4\pi)^2\gamma_0^{\text{stag},m} =\begin{pmatrix}
 K \\
 K \\
 L \\
4O \\
12L\\
 O \\
3L 
\end{pmatrix}.
\end{align}
Here we split the operators $\opSix_{11-38}$ into groups of four and introduced the following short-hands
\begingroup\allowdisplaybreaks
\begin{align}
A&=\left(\begin{smallmatrix}
 2 \hat{b}-6 \Nc+\frac{6}{\Nc} & 0 & 0 & 0 \\
 0 & 2 \hat{b}-6 \Nc+\frac{6}{\Nc} & 0 & 0 \\
 0 & 0 & 2 \hat{b}+\frac{6}{\Nc} & 0 \\
 0 & 0 & 0 & 2 \hat{b}+\frac{6}{\Nc}
\end{smallmatrix}\right),\quad
B=\left(\begin{smallmatrix}
 0 & 0 & 2 & 0\vphantom{\frac{8}{\Nc}} \\
 0 & 0 & 0 & 2\vphantom{\frac{8}{\Nc}} \\
 \frac{1}{2}-\frac{1}{2 \Nc^2} & 0 & \frac{2}{\Nc}-\frac{\Nc}{2} & 0 \\
 0 & \frac{1}{2}-\frac{1}{2 \Nc^2} & 0 & \frac{2}{\Nc}-\frac{\Nc}{2}
\end{smallmatrix}\right),\nonumber\\
C&=\left(\begin{smallmatrix}
 2 \hat{b} & 0 & 0 & -12\vphantom{\frac{8}{\Nc}} \\
 0 & 2 \hat{b} & -12 & 0\vphantom{\frac{8}{\Nc}} \\
 0 & \frac{3}{\Nc^2}-3 & 2 \hat{b}-3 \Nc & 3 \Nc-\frac{12}{\Nc} \\
 \frac{3}{\Nc^2}-3 & 0 & 3 \Nc-\frac{12}{\Nc} & 2 \hat{b}-3 \Nc
\end{smallmatrix}\right),\quad
D=\left(\begin{smallmatrix}
 2 \hat{b}+2 \Nc-\frac{2}{\Nc} & 0 & 0 & 0 \\
 0 & 2 \hat{b}+2 \Nc-\frac{2}{\Nc} & 0 & 0 \\
 0 & 0 & 2 \hat{b}-4 \Nc-\frac{2}{\Nc} & 0 \\
 0 & 0 & 0 & 2 \hat{b}-4 \Nc-\frac{2}{\Nc}
\end{smallmatrix}\right)\nonumber\\
E&=\left(\begin{smallmatrix}
 0 & 0 & 0 & 0\vphantom{\frac{8}{\Nc}} \\
 0 & 0 & 0 & 0\vphantom{\frac{8}{\Nc}} \\
 0 & 0 & -\Nc & 0\vphantom{\frac{8}{\Nc}} \\
 0 & 0 & 0 & -\Nc\vphantom{\frac{8}{\Nc}}
\end{smallmatrix}\right),\quad
F=\left(\begin{smallmatrix}
 2 \hat{b}+2 \Nc-\frac{2}{\Nc} & 0 & 0 & 0 \\
 0 & 2 \hat{b}+2 \Nc-\frac{2}{\Nc} & 0 & 0 \\
 0 & 0 & 2 \hat{b}-\frac{2}{\Nc} & 0 \\
 0 & 0 & 0 & 2 \hat{b}-\frac{2}{\Nc}
\end{smallmatrix}\right),\nonumber\\
G&=\left(\begin{smallmatrix}
 0 & 0 & 0 & -4\vphantom{\frac{8}{\Nc}} \\
 0 & 0 & -4 & 0\vphantom{\frac{8}{\Nc}} \\
 0 & \frac{1}{\Nc^2}-1 & -\Nc & \Nc-\frac{4}{\Nc} \\
 \frac{1}{\Nc^2}-1 & 0 & \Nc-\frac{4}{\Nc} & -\Nc
\end{smallmatrix}\right),\quad
H=\left(\begin{smallmatrix}
 2 \hat{b} & 0 & 0 & 4\vphantom{\frac{8}{\Nc}} \\
 0 & 2 \hat{b} & 4 & 0\vphantom{\frac{8}{\Nc}} \\
 0 & 1-\frac{1}{\Nc^2} & 2 \hat{b}+\Nc & \frac{4}{\Nc}-\Nc \\
 1-\frac{1}{\Nc^2} & 0 & \frac{4}{\Nc}-\Nc & 2 \hat{b}+\Nc
\end{smallmatrix}\right),\nonumber\\
I&=\left(\begin{smallmatrix}
 0 & 0 & 0 & 0 & 0 & \frac{16}{3} & 0 \\
 0 & 0 & 0 & 0 & 0 & -\frac{16}{3} & 0 \\
 0 & 0 & 0 & 0 & 0 & \frac{8}{3 \Nc} & 0 \\
 0 & 0 & 0 & 0 & 0 & -\frac{8}{3 \Nc} & 0
\end{smallmatrix}\right),\quad
J=\left(\begin{smallmatrix}
 0 & 0 & 0 & 0 & 0 & -\frac{8}{3} & 0 \\
 0 & 0 & 0 & 0 & 0 & -\frac{8}{3} & 0 \\
 0 & 0 & 0 & 0 & 0 & -\frac{4}{3 \Nc} & 0 \\
 0 & 0 & 0 & 0 & 0 & -\frac{4}{3 \Nc} & 0
\end{smallmatrix}\right),\quad
K=\left(\begin{smallmatrix}
 128 & 0 & 16 & 0\vphantom{\frac{8}{\Nc}} \\
 -128 & 0 & -16 & 0\vphantom{\frac{8}{\Nc}} \\
 \frac{64}{\Nc} & 0 & \frac{8}{\Nc}-8 \Nc & 0 \\
 -\frac{64}{\Nc} & 0 & 8 \Nc-\frac{8}{\Nc} & 0
\end{smallmatrix}\right),\nonumber\\
L&=\left(\begin{smallmatrix}
 0 & 0 & 16 & 0\vphantom{\frac{8}{\Nc}} \\
 0 & 0 & -16 & 0\vphantom{\frac{8}{\Nc}} \\
 0 & 0 & \frac{8}{\Nc}-8 \Nc & 0 \\
 0 & 0 & 8 \Nc-\frac{8}{\Nc} & 0 
\end{smallmatrix}\right),\quad
O=\left(\begin{smallmatrix}
 -128 & 0 & 48 & 0\vphantom{\frac{8}{\Nc}} \\
 128 & 0 & -48 & 0\vphantom{\frac{8}{\Nc}} \\
 -\frac{64}{\Nc} & 0 & \frac{24}{\Nc}-24 \Nc & 0 \\
 \frac{64}{\Nc} & 0 & 24\Nc-\frac{24}{\Nc} & 0
\end{smallmatrix}\right),\quad \hat{b}=(4\pi)^2b_0\,.
\end{align}
\endgroup
We only give those new anomalous dimension matrices for completeness.
Eventually we are interested in finding a choice of basis $\op\rightarrow\base$ that diagonalises the full 1-loop anomalous dimension matrix or at least turns it into Jordan normal form, which then allows us to write
\begin{equation}
\base_{i;\MSbar}(\mu)=[2b_0\gbar^2(\mu)]^{\hat{\gamma}_i}\exp\left[\left\{\frac{\gamma_0^\base}{2b_0}-\hat{\gamma}\right\}\log\left(2b_0\gbar^2(\mu)\right)\right]_{ij}\base_{j;\text{RGI}}\times\{1+\ord(\gbar^2(\mu))\}.\label{eq:connectionRGI}
\end{equation}
Here $\gamma_0^\base$ is in Jordan normal form and $\base_{i;\text{RGI}}$ is a Renormalisation Group Invariant constant and all (leading order) scale-dependence has now been absorbed into the prefactor, while we introduced the short-hand
\begin{equation}
\hat{\gamma}=\diag\left(\frac{\gamma_0^\base}{2b_0}\right).
\end{equation}
Those $\hat{\gamma}_{i}$ are precisely the powers in the running coupling we set out to compute.
We give their numerical values in \tab{tab:gammahat} for the four viable options of $\Nf=0,4,8,12$ without loosing asymptotic freedom entirely.\footnote{Putting the question aside whether QCD becomes (near-)conformal at $\Nf=12$ for $\Nc=3$.}
In case the 1-loop anomalous dimension matrix is non-diagonalisable, the argument of the exponential in \eq{eq:connectionRGI} does not vanish and we find explicit factors of $\log\left(2b_0\gbar^2(\mu)\right)$ multiplying some of the overall $[2b_0\gbar^2(\mu)]^{\hat{\gamma_i}}$ powers to leading order.
For a more in-depth discussion of this, see \cite{Husung:2022kvi}.
Here this complication only arises for the quenched case, i.e., $\Nf=0$.

\begin{table}\centering
\caption{Overview of the first five unique asymptotically leading powers $\hat{\gamma}_i$ found at $\ord(a^2)$ \emph{without} taking suppression due to vanishing matching coefficients or any subleading corrections into account.
The latter will obviously start at $\hat{\gamma}_i+1$.
\dotuline{Underdotted} numbers belong to explicitly mass-dependent contributions, \underline{underlined} numbers are compatible with chiral symmetry, and the \textbf{bold} number originates from both the chirally symmetric basis and the Staggered extension where the latter also gives rise to an explicit $[2b_0\gbar^2(1/a)]^{0.273}\log(2b_0\gbar^2(1/a))$ contribution.}
\label{tab:gammahat}
\begin{tabular}{c|l}
$\Nf$ & $\hat{\gamma}_i$ \\\hline
0     & \dotuline{$-0.273$}, $0.014$, {\boldmath $0.273$}, \dotuline{$0.424$}, $0.560$, ...\\
4     & $-0.301$, \dotuline{$-0.040$}, $0.040$, \underline{$0.210$}, $0.419$, ...\\
8     & $-0.913$, $-0.412$, \underline{$-0.103$}, $0.146$, $0.294$, ...\\
12    & $-2.614$, $-1.667$, \underline{$-1.040$}, $-0.614$, $-0.333$, ...
\end{tabular}
\end{table}

%% file: concl.tex
\section{Discussion}
Independently from, e.g.,~\cite{Luo:1996vt,Lee:1999zxa,Follana:2006rc} we worked out the minimal on-shell operator basis for the SymEFT action of unrooted Staggered quarks.
We find the same number of taste-breaking 4-quark operators as has been previously found in~\cite{Follana:2006rc}.
However, both results disagree with \cite{Lee:1999zxa} and earlier works.
Indeed our 1-loop renormalisation requires all of the 4-quark operators listed here as counterterms.\footnote{Conceptually this only guarantees that our basis is minimal in the sense of absence of linear dependencies but it may still be incomplete.
The latter seems unlikely given that no counterterms are missing.}

This minimal on-shell basis is sufficient for describing the asymptotically leading lattice artifacts of spectral quantities such as hadron masses.
We computed the corresponding powers $a^2[2b_0\gbar^2(1/a)]^{\hat{\gamma}_i}$, where we set $\mu=1/a$ as the relevant renormalisation scale of lattice artifacts.
For $\Nf=0,4$ we find $\min_i\hat{\gamma}_i\approx -0.3$ which lies well within previous results found for Wilson and GW quarks and is much better behaving than $\min_i\hat{\gamma}_i\approx -3$ as found in the past for the O(3) model~\cite{Balog:2009yj,Balog:2009np}.
However, for $\Nf=8$ and $\Nf=12$ we find increasingly negative powers $\min_i\hat{\gamma}_i\approx -0.913$ and $\min_i\hat{\gamma}_i\approx -2.614$ respectively.
This hints at a severe worsening of the naively expected $a^2$-behaviour, i.e., taking the continuum limit may become much more complicated than it already is for conventional lattice QCD.
Keep in mind that those powers arise from chiral-symmetry breaking 4-quark operators which are expected to be tree-level suppressed when performing the matching to the lattice theory.
Hence those powers of concern are expected to be shifted by $+1$, which is very similar to our results for the chiral-symmetry breaking 4-quark sector of Wilson lattice QCD~\cite{Husung:2022kvi}.

As already stressed in previous works, having the renomalisation scale $\mu=1/a$ brings us only close to the beginning of the perturbative regime in nowadays lattice setups, where $\alpha_{\MSbar}(1/a)=\gbar^2(1/a)/(4\pi)\sim 0.2$.\footnote{Keep in mind that Staggered quarks are formulated on a coarser lattice with lattice spacing $2a$ and it is unclear if not $\mu=1/(2a)$ is a more appropriate choice.}
Thus working at leading logarithmic order may be insufficient.
From that viewpoint and owing to the fact that $\ord(a^3)$ corrections and above may still be relevant, reaching smaller lattice spacings remains highly desirable.

This work also clarifies the field-theoretical argument for the Shift-symmetry constraints on the minimal on-shell basis in a taste-representation strictly local in spacetime.
Conceptually, we obtain two minimal bases, namely an on-shell basis and an EOM-vanishing basis respectively.
The latter has severely relaxed symmetry constraints but only the subset of operators compatible with the symmetries of the minimal on-shell basis is needed here to perform the 1-loop renormalisation off-shell.
Meanwhile, the EOM-vanishing operators of mass-dimension~5 in \eq{eq:opFive} will need to be taken into account when performing the tree-level matching of the on-shell SymEFT action to $\ord(a^2)$.

Also any Symanzik analysis (or improvement) of local composite fields will be impacted by the full EOM-vanishing basis analogously to the analysis of local fermion bilinears for Wilson or GW quarks~\cite{Capitani:1999ay,Capitani:2000xi,Husung:localFields}.
In particular, absence of $\ord(a)$ lattice artifacts is only guaranteed by automatic $\ord(a)$ improvement due to the discrete symmetry in \eq{eq:tau5Symmetry} of the continuum theory.
For the improvement of current-current 2-point functions this is a relevant contribution already at tree-level and $\ord(a^2)$.
The full basis of EOM-vanishing operators at $\ord(a^2)$ may be significantly enlarged due to relaxed symmetry constraints from rotations and Euclidean reflections.

In light of this, a full-fledged Symanzik analysis for local mass-dimension~3 bilinears of unrooted Staggered quarks seems desirable, analogously to Wilson and GW quarks~\cite{Husung:localFields}.
It would further be interesting to extend the SymEFT analysis within perturbation theory to rooted Staggered quarks~\cite{Gottlieb:1987mq} following along the lines of rooted Staggered chiral perturbation theory~(s$\chi$pt), see e.g.~\cite{Bernard:2001yj,Aubin:2003rg,Bernard:2006zw}.
From a SymEFT point of view as presented here, the non-locality of the rooted theory at finite lattice spacing remains an issue.
Questions to be understood are the proper order of limits for rooting and the continuum limit as has been discussed in the context of s$\chi$pt, see e.g.~\cite{Bernard:2004ab}.

\paragraph{Supplementary material.}
A rudimentary \texttt{Mathematica} notebook (\texttt{symm1CRtoTastes.nb}) is supplied alongside this manuscript that can be used to derive the symmetry transformations in the local TR used here starting from the 1CR.
Another \texttt{Mathematica} file (\texttt{staggered.wl}) is supplied as well to allow direct access to the 1-loop anomalous dimension matrix derived in this paper including the block matrices omitted here.
Building on previous work, \texttt{FORM}~\cite{Vermaseren:2000nd} scripts, Python scripts and \texttt{Mathematica} notebooks including some automation via a \texttt{Makefile} have been adapted to Staggered quarks in a separate branch of the original git-repository, which is publicly available.\footnote{\url{https://github.com/nikolai-husung/Symanzik-QCD-workflow/tree/Staggered}}

\paragraph{Acknowledgements.}
I thank Carlos Pena, Gregorio Herdoíza, and Rainer Sommer for encouraging me to extend the SymEFT analysis to Staggered quarks.
I also thank Stefan Sint, Oliver B\"ar, Maarten Golterman, and Agostino Patella for discussions regarding the correct choice of symmetry constraints providing valuable insight into relating the non-trivial lattice and continuum symmetry transformations.
I am indebted to Gregorio Herdoíza for very helpful suggestions and discussions on the project and to Mattia Dalla Brida for comments on the manuscript.
Finally I thank the IFT, DESY Zeuthen, the Humboldt-University, and CERN for providing a hospitable environment for all these discussions to take place.

The author acknowledges funding by the STFC consolidated grant ST/T000775/1 as well as support of the projects PID2021-127526NB-I00, funded by MCIN/\allowbreak AEI/\allowbreak 10.13039/\allowbreak 501100011033 and by FEDER EU, IFT Centro de Excelencia Severo Ochoa No CEX2020-001007-S, funded by MCIN/\allowbreak AEI/\allowbreak 10.13039/\allowbreak 501100011033, H2020-MSCAITN-2018-813942 (EuroPLEx), under grant agreement No. 813942, and the EU Horizon 2020 research and innovation programme, STRONG-2020 project, under grant agreement No. 824093.

%% file: appa.tex
\section{List of (here) relevant EOM-vanishing operators}\label{sec:EOMops}
In order to perform the off-shell renormalisation of the minimal on-shell basis, we need to keep track of the minimal basis of EOM-vanishing operators compatible with all the symmetries of the on-shell operator basis.
While those mix under renormalisation with our on-shell basis, they do not affect spectral quantities.
From the analysis of GW quarks~\cite{Husung:2022kvi}, we already know the set of EOM-vanishing operators compatible with (spurionic) chiral symmetry
\begin{align}
\opSixE[;1] &= \frac{1}{g_0^2}\tr([D_\mu, F_{\mu\rho}] [D_\nu, F_{\nu\rho}])+\frac{1}{2}\bar\Psi \gamma_\mu\otimes\unity  [D_\nu, F_{\nu\mu}]\Psi, &
\opSixE[;2] &= \frac{1}{2}\bar\Psi\big\{D^2\Dslash-\cev{\Dslash}\cev{D}^2\big\}\Psi, \nonumber\\
\opSixE[;3] &= \bar\Psi \gamma_\mu\otimes\unity  [D_\nu, F_{\nu\mu}]\Psi-g_0^2(\bar\Psi\gamma_\mu\otimes\unity  T^a\Psi)^2, &
\opSixE[;4] &= \bar\Psi \Dslash^3\Psi, \nonumber\\
\opSixE[;5] &= \bar\Psi M \Dslash^2 \Psi, &
\opSixE[;6] &= \bar\Psi M^2 \Dslash \Psi, \nonumber\\
\opSixE[;7] &= \tr(M^2) \bar\Psi \Dslash \Psi.\label{eq:EOMa2Action}
\end{align}
We omit any other EOM-vanishing operators allowed that violate the canonical form of \emph{modified} discrete rotations, \emph{modified} Euclidean reflections or Shift-symmetry due to the presence of field redefinitions.
Those additional operators play no role for the renormalisation of the basis for the SymEFT action but may become relevant for the Symanzik description of local composite fields.

\section{Symmetries in the 1CR}\label{sec:1CRsymmetries}
Those are the transformations in the 1CR keeping the action from \eq{eq:staggeredAction} invariant
\begingroup\allowdisplaybreaks
\begin{align}
&\mathrlap{\text{\textbf{\boldmath``Eucl. reflection'' in direction $\hat{\mu}$ with a shift}}}\nonumber\\*
&\bar\chi(x)\rightarrow (-1)^{x_\mu/a}\zeta_\mu(x)\bar\chi(x-2x_\mu\hat{\mu}+a\hat{\mu})U^\dagger(x-2x_\mu\hat{\mu},\mu),\nonumber\\*
&\chi(x)\rightarrow (-1)^{x_\mu/a}\zeta_\mu(x)U(x-2x_\mu\hat{\mu},\mu)\chi(x-2x_\mu\hat{\mu}+a\hat{\mu}),\nonumber\\ &U(x,\nu)\rightarrow\begin{cases}
U^\dagger(x-(2x_\mu+a)\hat{\mu},\mu) & \mu=\nu \\
U(x-2x_\mu \hat{\mu},\mu)U(x-(2x_\mu-a)\hat{\mu},\nu)U^\dagger(x-2x_\mu\hat{\mu}+a\hat{\nu},\mu) & \text{else}
\end{cases}\\
\mathrlap{\text{\textbf{\boldmath``Charge conjugation''}}}\nonumber\\
&\bar\chi(x)\rightarrow e^{i\pi\eta_5(x)/2}\chi(x),\quad\chi(x)\rightarrow e^{i\pi\eta_5(x)/2}\bar\chi(x),\quad U(x,\mu)\rightarrow U^*(x,\mu)\\
\mathrlap{\text{\textbf{\boldmath``Rotation'' in $(\rho,\sigma)$-plane $\rho\gtrless\sigma$ with a shift}}}\nonumber\\*
&\bar\chi(x)\rightarrow S_{\rho\sigma}(R^{-1}x)\zeta_\sigma(R^{-1}x)\bar\chi(R^{-1}x+a\hat{\sigma})U^\dagger(R^{-1}x,\sigma),\nonumber\\*
&\chi(x)\rightarrow S_{\rho\sigma}(R^{-1}x)\zeta_\sigma(R^{-1}x)U(R^{-1}x,\sigma)\chi(R^{-1}x+a\hat{\sigma}),\nonumber\\*
&U(x,\nu)\rightarrow \begin{cases}
U^\dagger(R^{-1}x-a\hat{\sigma},\sigma) & \nu=\rho\\
U(R^{-1}x,\sigma)U(R^{-1}x+a\hat{\sigma},\rho)U^\dagger(R^{-1}x+a\hat{\rho},\sigma) & \nu=\sigma\\
U(R^{-1}x,\sigma)U(R^{-1}x+a\hat{\sigma},\nu)U^\dagger(R^{-1}x+a\hat{\nu},\sigma) & \text{else}
\end{cases}\nonumber\\*
&S_{\rho\sigma}(x)=\frac{1}{2}\left[1\pm\eta_\rho(x)\eta_\sigma(x)\mp\zeta_\rho(x)\zeta_\sigma(x)+\eta_\rho(x)\eta_\sigma(x)\zeta_\rho(x)\zeta_\sigma(x)\right],\nonumber\\*
&(R^{-1}x)_{\nu\neq\rho,\sigma}=x_\nu,\quad (R^{-1}x)_\rho= x_\sigma,\quad (R^{-1}x)_\sigma= -x_\rho\\
\mathrlap{\text{\textbf{\boldmath``Shift'' in direction $\hat{\mu}$}}}\nonumber\\*
&\bar\chi(x)\rightarrow\zeta_\mu(x)\bar\chi(x+a\hat{\mu})U^\dagger(x,\mu),\quad \chi(x)\rightarrow\zeta_\mu(x)U(x,\mu)\chi(x+a\hat{\mu}),\nonumber\\*
&U(x,\nu)\rightarrow U(x,\mu)U(x+a\hat\mu,\nu)U^\dagger(x+a\hat{\nu},\mu),\label{eq:1CRshift}\\
\mathrlap{\text{\textbf{``Remnant chiral symmetry''}}}\nonumber\\*
&\bar\chi(x)\rightarrow e^{i\vartheta\eta_5(x)-i\varphi}\bar\chi(x),\quad \chi(x)\rightarrow e^{i\vartheta\eta_5(x)+i\varphi}\chi(x),\quad \varphi,\vartheta\in\mathbb{R},
\end{align}
where we introduced the shorthand commonly found in the literature
\begin{equation}
\zeta_\mu(x)=(-1)^{\sum_{\nu>\mu}x_\nu/a}\,.
\end{equation}
\endgroup
The phase introduced in~\cite{Verstegen:1985kt} into what we refer to as ``charge conjugation'' is necessary to ensure that in the TR this corresponds to \emph{modified} charge conjugation rather than a combination of \emph{modified} charge, and remnant chiral symmetry.
Reflections and rotations move the hypercube and we thus combine them with appropriate shifts to return to the initial grouping.
For convenience we introduced gauge links to keep the reference point fixed when performing a shift.
Eventually, this amounts to a gauge transformation and thus could be omitted but it helps to guide at which lattice point one currently is when switching hence and forth between 1CR and TR.

The symmetry transformations in the 1CR are mostly identical to~\cite{Verstegen:1985kt} but we work in the interacting theory.
Those symmetry transformations are very similar to the ones found in~\cite{Golterman:1984cy} but already incorporate the shifts necessary to return to the proper hypercube.
Somewhat puzzling, the rotation matrix $S_{\rho\sigma}(x)$ is different between both publications and from the one we need to use here to keep the 1CR action invariant.
We find opposite signs for $\rho\gtrless\sigma$ compared to \cite{Golterman:1984cy} despite relying on the same equation~(2.8) therein while \cite{Verstegen:1985kt} does not even differentiate between both cases.

\section{Example: Modified Euclidean reflection in $\hat{1}$-direction in the local taste representation}\label{sec:RedefExample}
When working out the actual symmetries in the local TR one finds rather complicated transformations that involve various Wilson lines and Dirac matrices acting in spinor and/or taste space.
This is due to the fact that after any transformation the initial choice for defining tastes from \eq{eq:spacetimeTastes} has to be recovered.
In most cases a linear combination of different Wilson lines multiplied by various spinor and taste-space matrices is required to achieve this.
Such transformations are no longer conventional symmetry transformations but involve an additional field-redefinition.

Due to its relatively compact form in our chosen TR, let us consider the \emph{modified} Euclidean reflection in $\hat{1}$-direction as an example
\begin{align}
\tastebar(y-2y_\mu\hat{\mu}) &\rightarrow \frac{1}{2}\tastebar(y)\gamma_5\gamma_1\otimes\tau_5\Big\{1+\mathcal{W}^\dagger(y)+\gamma_0\gamma_5\otimes\tau_0\tau_5\left(\mathcal{W}^\dagger(y)-1\right)\Big\},\nonumber\\
\taste(y-2y_\mu\hat{\mu}) &\rightarrow \frac{1}{2}\Big\{1+\mathcal{W}(y)+\gamma_0\gamma_5\otimes\tau_0\tau_5\left(\vphantom{\mathcal{W}^\dagger}\mathcal{W}(y)-1\right)\Big\}\gamma_1\gamma_5\otimes\tau_5\taste(y),\nonumber\\
U(x-2x_\mu\hat{\mu},\nu) &\rightarrow \begin{cases}
U^\dagger(x-a\hat{\mu},\mu) & \mu=\nu \\
U(x,\mu)U(x+a\hat{\mu},\nu)U^\dagger(x+a\hat{\nu},\mu) & \text{else}
\end{cases}\nonumber\\
\mathcal{W}(x)&=U(x,1)U(x+a\hat{1},0)U^\dagger(x+a\hat{0},1)U^\dagger(x,0) \stackrel{\wedge}{=} \mathbb{1} + a^2 \hat{F}_{10}(x)\,,
\end{align}
where we identify $\hat{F}_{10}$ as a crude lattice discretisation of the field-strength tensor.
This only serves as a reminder why we expect an impact at $\ord(a^2)$ at the earliest.
In the free theory the gauge links become the identity and only the conventional symmetry transformation remains.
Contrarily, in the fully interacting theory the transformation can be split into two parts.
Firstly the actual symmetry transformation that one would expect in continuum field theory and secondly a field-redefinition.
This split has already been implemented here by proper factorisation.
It should be clear that the Jacobians from the field-redefinitions of both $\tastebar$ and $\taste$ have to cancel exactly, which is indeed the case here.

\section{Restoring symmetries involving field-redefinitions to their canonical form}\label{sec:symmEOMrestoration}
The starting point of this discussion is the generic symmetry transformation of a field $\phi$
\begin{equation}
\phi\rightarrow \left\{1+a^d\square^{(d)}[\phi]+\ord(a^{d+1})\right\}\mathcal{T}\phi,
\end{equation}
where $\square^{(d)}[\phi]$ is a monomial of fields, derivatives etc.
This transformation keeps the Effective Lagrangian
\begin{equation}
\Leff[\phi]=\L[\phi]+\sum_{n\geq 1}a^n\L^{(n)}[\phi]
\end{equation}
invariant, where $\L[\phi]$ is the continuum Lagrangian and $\L^{(n)}[\phi]$ are the lattice-corrections at $\ord(a^{n})$.
For $n<d$ the symmetry of the lattice theory in their CCF, i.e., $\phi\rightarrow \mathcal{T}\phi$, keeps the action (to this mass-dimension) invariant while $n\geq d$ requires the field-redefinition to keep the action invariant.
In other words
\begin{equation}
\L^{(d)}[\phi]\rightarrow \L^{(d)}[\phi]-a^d\left\{\partial_\mu\frac{\delta\L[\phi]}{\delta\partial_\mu\phi}-\frac{\delta\L[\phi]}{\delta\phi}\right\}\square^{(d)}[\phi]\phi+\ord(a^{d+1})
\end{equation}
under the CCF symmetry transformation.
Thus the leading order part incompatible with the CCF of the symmetry transformation must be vanishing by the continuum EOMs.
Removing this part entirely at $\ord(a^{d})$ through an appropriate field-redefinition (or more naturally a change of matching condition), then amounts to $d\rightarrow d+1$ in the previous argument.
Consequently the $\ord(a^{d+1})$ Lagrangian is now the first order requiring a field-redefinition.
Eventually we can restore the CCF of the symmetry transformation order by order in the lattice spacing.
The derivation used here is agnostic to the EFT being a SymEFT and should be applicable to any other EFT although broad applicability seems unlikely.

%% file: SymanzikStaggered.bbl
\providecommand{\href}[2]{#2}\begingroup\raggedright\endgroup